# Spatiotemporal Diffusion Model with Paired Sampling for Accelerated Cardiac Cine MRI

*Shihan Qiu[1,2,3]\*, Shaoyan Pan[1,4,5]\*, Yikang Liu[1], Lin Zhao[1], Jian Xu[6], Qi Liu[6], Terrence Chen[1], Eric Z. Chen[1], Xiao Chen[1], Shanhui Sun[1]*

[1]United Imaging Intelligence, Burlington, Massachusetts, USA
[2]Biomedical Imaging Research Institute, Cedars-Sinai Medical Center, Los Angeles, California, USA
[3]Department of Bioengineering, UCLA, Los Angeles, California, USA
[4]Department of Radiation Oncology and Winship Cancer Institute, Emory University, Atlanta, GA, USA
[5]Department of Biomedical Informatics, Emory University, Atlanta, GA, USA
[6]UIH America, Inc., Houston, TX, USA

**Synopsis** (100-word)
Motivation: Current deep learning reconstruction for accelerated cardiac cine MRI suffers from spatial and temporal blurring.
Goals: To improve image sharpness and motion delineation for cine MRI under high undersampling rates.
Approach: A spatiotemporal diffusion enhancement model conditional on an existing deep learning reconstruction along with a novel paired sampling strategy was developed.
Results: The diffusion model provided sharper tissue boundaries and clearer motion than the original reconstruction in experts' evaluation on clinical data. The innovative paired sampling strategy substantially reduced artificial noises in the generative results.

**Impact** (40 words)
The approach has the potential to improve reconstruction quality in highly accelerated cardiac cine imaging. The novel paired sampling for diffusion generation may be applied to other conditional tasks to reduce the artificial noises stemming from noisy training data.

**Introduction**
Accelerated cardiac cine imaging such as real-time cine is highly desirable given its shortened scan time and increased motion robustness, but the reconstruction is challenging due to high undersampling rates. Although deep learning (DL) methods have significantly reduced aliasing artifacts in highly-accelerated MRI reconstruction, most of them formulate the reconstruction as a non-generative (regression) problem, resulting in compromised sharpness and motion blurring.[1,2] Diffusion models[3,4] as generative models, however, provide results by sampling from a learned distribution, and have shown promising results in narrowing the gap to fully-sampled quality in static image reconstruction but not for dynamic imaging.[5-9]
In this study, a combined non-generative DL reconstruction and diffusion generation framework was proposed for cine CMR. The diffusion part utilized spatiotemporal architectures to handle the 2D+time data and a novel paired-sampling strategy was additionally proposed to reduce the synthetic noises in the generative results stemming from noisy training data. The approach was evaluated quantitatively and qualitatively on both retro- and pro-spectively undersampled cine imaging. To our knowledge, this is the first study to reconstruct dynamic CMR images with diffusion models.

**Methods**
Combination of non-generative and generative models
A residual convolutional recurrent neural network (res-CRNN)[1] trained with paired under- and fully-sampled images was first used for initial reconstruction from undersampled raw data. The initial de-aliased result was fed to a diffusion model that was conditioned on the collected data and the de-aliased result (Figure 1). The diffusion model was trained with pairs of res-CRNN outputs and fully-sampled images. To handle dynamic data, a spatiotemporal network was adopted as the diffusion model to utilize the temporal information, which was constructed by adding 3D spatiotemporal convolution and temporal attention layers to a 2D U-net architecture.[10]

Paired sampling
In medical images, noises on reference fully sampled images are arguably inevitable which can be picked up by generative models to produce undesired artificial noises. We proposed a novel paired sampling strategy to address this issue. In a conditional generation task, the result **x** was hypothesized to be $\mathbf{x} = \mathbf{x}_s + \mathbf{x}_n$, where $\mathbf{x}_n$ represents the noises and $\mathbf{x}_s$ is the desired structure. We further assumed $\mathbf{x}_s$ to be from an interaction between the given condition and the noises input to the diffusion system, whereas $\mathbf{x}_n$ mainly calculated from the noises themselves. A paired sampling was designed, where opposite noises $\mathbf{z}_{pos}$ and $\mathbf{z}_{neg}$ were inputted for two samplings, resulting in $\mathbf{x}_{pos} = \mathbf{x}_{pos,s} + \mathbf{x}_{pos,n}$, and $\mathbf{x}_{neg} = \mathbf{x}_{neg,s} + \mathbf{x}_{neg,n}$. It was oberserved that the synthetic noises in these samplings were also largely opposite to each other, indicating that $\mathbf{x}_{pos,n} \approx -\mathbf{x}_{neg,n}$. Additionally, $\mathbf{x}_{pos,s} \approx \mathbf{x}_{neg,s} \approx \mathbf{x}_s$ according to the definition. Therefore, an average of the two was output as the final result: $\mathbf{x}_{pair} = (\mathbf{x}_{pos} + \mathbf{x}_{neg})/2 \approx \mathbf{x}_s$.

Data and experiments
The approach was first trained and evaluated on bSSFP retro-cine data collected on 3T MRI scanners (uMR790, UIH, Shanghai, China) with IRB approval, where retrospective undersampling (R=8~16) was applied to 1071 slices from 43 subjects (6/2/2 for train/validation/test splitting). Clinical real-time cine from 2 subjects were additionally acquired for testing. A spatial resolution of 1.82x1.82 mm² and a temporal resolution of 34 ms and 42 ms for retro and real-time, respectively, were used. Quantitative metrics and qualitative blind rankings from two experienced (>10yr) experts were used for quality assessment.

**Results**
The combined res-CRNN and diffusion framework achieved a similar image appearance to fully sampled images, with sharper tissue boundaries and reduced temporal blurring than the original res-CRNN reconstruction (Figure 2). The paired sampling strategy provided much lower noises compared to single sampling even with averages. The proposed method achieved the best NMSE, PSNR, and SSIM (Table 1). In the visual evaluation, our approach achieved better performance than res-CRNN reconstruction in all aspects (Table 2). Testing on real-time cine also showed consistent performance in reducing blurring and artifacts (Figure 3 & Table 2).

**Discussion**

The proposed method demonstrated the effectiveness of combining the synergy of non-generative reconstruction and diffusion generative models to reconstruct high-quality cine images from highly accelerated data. The paired sampling efficiently removed the undesired noise prior embedded in the training data, showing much higher PSNR, SSIM, and preferred visual appearance over the conventional diffusion methods.

**Conclusion**

A combined non-generative reconstruction and diffusion generation was developed to improve the image quality in highly accelerated cine reconstruction. The proposed model along with a paired sampling strategy provided effective enhancement of sharpness while mitigating the generation of artificial noises, resulting in high-quality cine images.

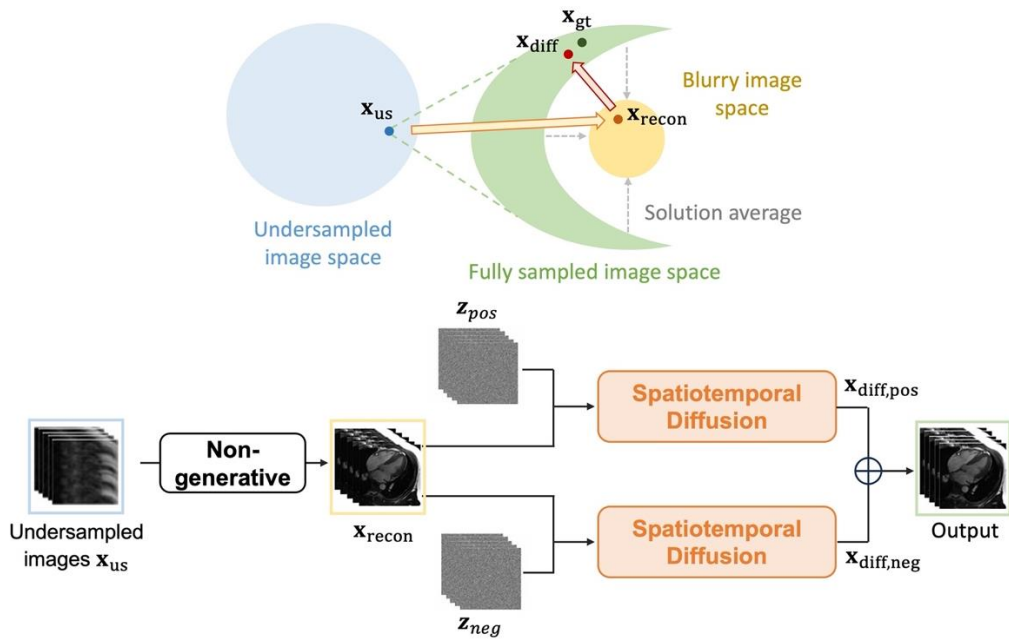

Figure 1. Illustration of the proposed method. A non-generative DL reconstruction was first applied to obtain an initial de-aliased reconstruction but with blurring due to solution average. Then a spatiotemporal diffusion model took this reconstruction for further enhancement to match fully sampled distribution and provide sharp images. This diffusion sampling was performed twice, with a pair of input noises, and the results were combined for the final output.

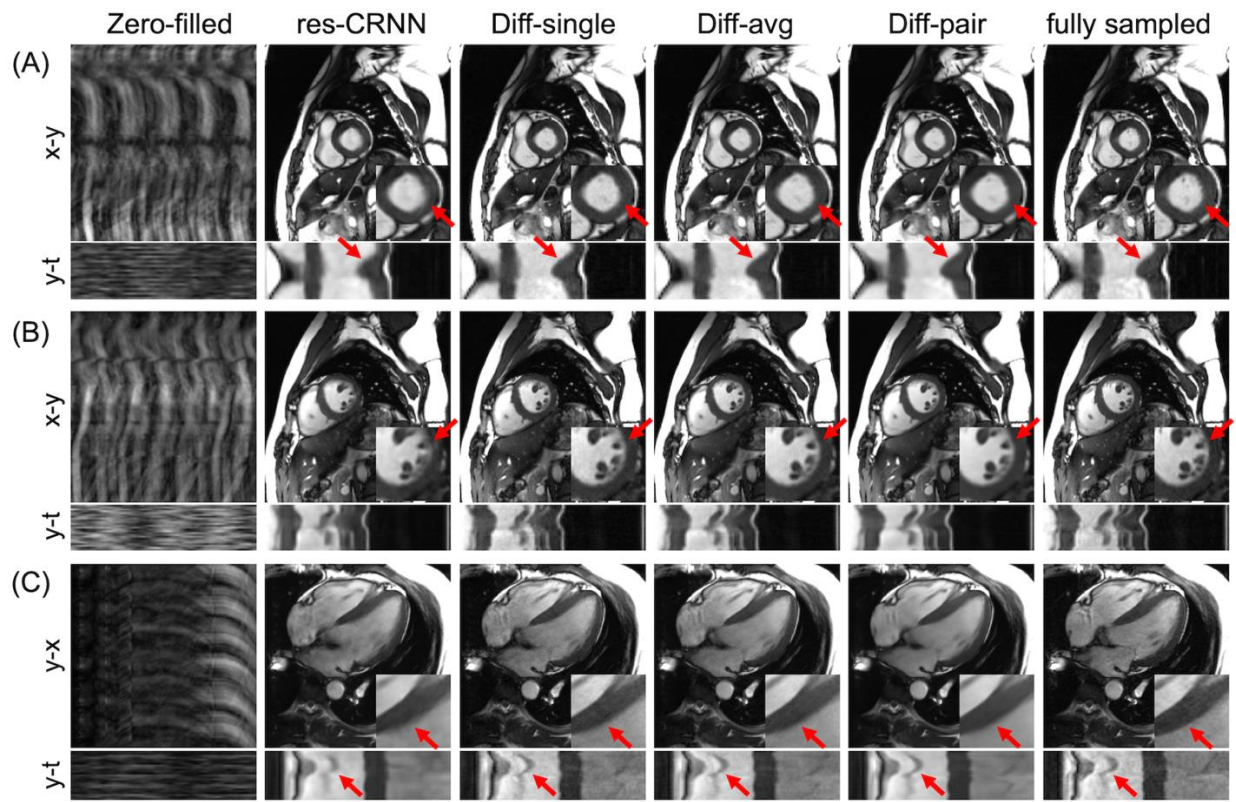

Figure 2. Results on three slices from retro-cine data. The diffusion methods reduced spatial and temporal blurring in res-CRNN reconstruction results (red arrows). The paired sampling strategy (Diff-pair) substantially removed the artificial noises generated by the diffusion model due to the noisy fully sampled images used in the training, with higher efficiency than a simple average on two random samplings (Diff-avg).

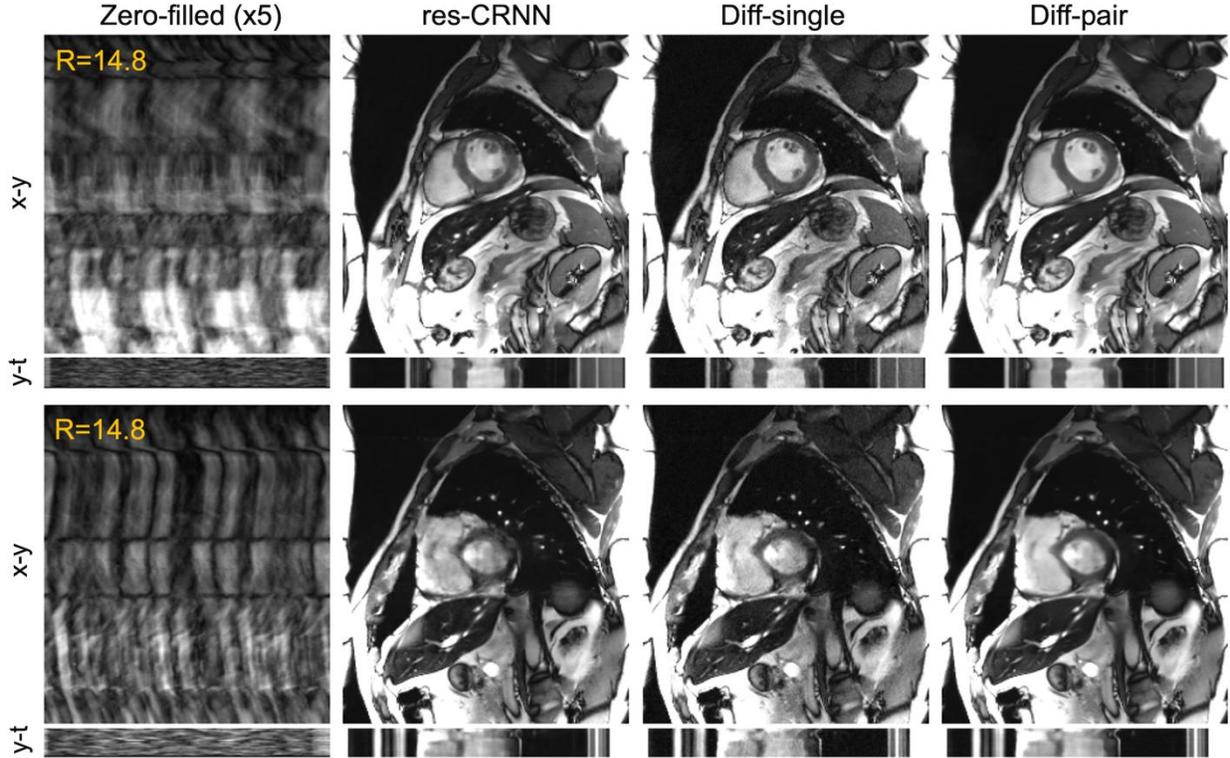

Figure 3. Example images from two real-time cine slices. The proposed diffusion method (Diff-pair) improved the sharpness at tissue boundaries and mitigated remaining aliasing artifacts compared to the original res-CRNN reconstruction at a high undersampling rate (R=14.8). It also had much fewer noises than a single diffusion sampling.

Table 1. Quantitative evaluation. The asterisks indicate statistically significant differences (P<0.05) when compared with res-CRNN reconstruction. The proposed diffusion model with paired sampling (Diff-pair) achieved the best metrics. Single sampling (Diff-single) had lower PSNR and SSIM than res-CRNN reconstruction mainly due to the artificial noises in the images. Averaging on two random samplings (Diff-avg) mitigated these noises and improved the metrics, but was not as efficient as Diff-pair.

|  | NMSE | PSNR (dB) | SSIM |
| --- | --- | --- | --- |
| res-CRNN recon | 0.0027±0.0015 | 43.65±2.38 | 0.9837±0.0085 |
| Diff-single | 0.0036±0.0019* | 42.35±2.28* | 0.9778±0.0113* |
| Diff-avg | 0.0031±0.0016* | 43.05±2.31* | 0.9812±0.0097* |
| Diff-pair (proposed) | **0.0025±0.0014*** | **43.88±2.35*** | **0.9846±0.0080*** |

Table 2. Rankings from two experienced experts on 24 retro-cine slices and 10 real-time cine slices. Rank 1 is the best. Bold numbers indicate the best performance among all methods. The asterisks indicate statistically significant differences (P<0.05) when compared with res-CRNN reconstruction. The proposed method (Diff-pair) achieved better rankings than the original res-

---

*This work was carried out during the internship of the author at United Imaging Intelligence, Burlington, MA 01803

CRNN reconstruction in all aspects. It also performed better than the single sampling (Diff-single) as well as fully sampled images in noise and overall quality.

|  |  | res-CRNN recon | Diff-single | Diff-pair | fully sampled |
|---|---|---|---|---|---|
| Retro-cine | Sharpness | 3.9±0.3 | 1.9±0.5* | 2.9±0.6* | **1.2±0.5** |
|  | SNR | 1.9±0.4 | 3.0±0.1* | **1.1±0.3*** | 4.0±0.1 |
|  | Temporal Sharpness | 3.9±0.3 | 2.0±0.4* | 3.0±0.4* | **1.1±0.3** |
|  | Overall Quality | 2.9±0.9 | 2.2±0.9* | **1.9±1.1*** | 3.0±1.1 |
| Real-time cine | Sharpness | 3.0±0.0 | **1.3±0.5*** | 1.7±0.5* | -- |
|  | SNR | 2.3±0.5 | 2.7±0.5 | **1.0±0.0*** | -- |
|  | Temporal Sharpness | 3.0±0.2 | **1.2±0.4*** | 1.8±0.5* | -- |
|  | Overall Quality | 3.0±0.2 | 2.0±0.3* | **1.1±0.2*** | -- |